\pdfoutput=1
\documentclass[11pt]{article}

\usepackage[utf8]{inputenc}
\usepackage[T1]{fontenc}
\usepackage{lmodern}
\usepackage[margin=1in]{geometry}
\usepackage{amsmath,amssymb}
\usepackage{graphicx}
\usepackage{booktabs}
\usepackage{tabularx}
\usepackage{array}
\usepackage{enumitem}
\usepackage[numbers,sort&compress]{natbib}
\usepackage{xcolor}
\usepackage{titlesec}

\definecolor{KnittInk}{HTML}{342F28}
\definecolor{KnittBrown}{HTML}{8C6A3F}
\definecolor{KnittOchre}{HTML}{C9A24A}
\definecolor{KnittSand}{HTML}{F1EADD}
\definecolor{KnittMoss}{HTML}{697957}
\definecolor{KnittSlate}{HTML}{5B7580}
\definecolor{KnittMist}{HTML}{E9EEEE}
\definecolor{KnittPaper}{HTML}{FCFAF4}

\titleformat{\section}
  {\large\bfseries\color{KnittInk}}
  {\thesection}{0.65em}{}
  [\vspace{0.15em}{\color{KnittBrown!60}\titlerule[0.45pt]}]
\titleformat{\subsection}
  {\normalsize\bfseries\color{KnittMoss}}
  {\thesubsection}{0.6em}{}
\titleformat{\paragraph}[runin]
  {\normalsize\bfseries\color{KnittInk}}
  {}{0pt}{}[.]

\usepackage{tikz}
\usetikzlibrary{positioning,arrows.meta,fit,calc}
\usepackage{microtype}
\usepackage{hyperref}
\hypersetup{
  colorlinks=true,
  linkcolor=KnittSlate,
  citecolor=KnittMoss,
  urlcolor=KnittBrown
}
\usepackage[capitalise,noabbrev]{cleveref}

\setlist{itemsep=2pt,topsep=4pt}

\newcolumntype{L}{>{\raggedright\arraybackslash}X}
\newcommand{\code}[1]{\texttt{#1}}
\newcommand{\tk}{teamknitt}

\title{\textcolor{KnittInk}{\textbf{teamknitt: Configuring, Observing and Intervening\\in LLM-Agent Teams through the Lens of\\Dynamical-Systems and Human-Team Theory}}}
\author{Goran Mitevski\\[2pt]
  \small \href{mailto:rolevski.mk@icloud.com}{\texttt{rolevski.mk@icloud.com}}}
\date{November 2026}

\begin{document}
\color{KnittInk!96}
\maketitle

\begin{abstract}
Most frameworks for teams of large language model (LLM) agents fix the team's shape inside the
software: who talks to whom, how much is passed along, and what a human can do once a run has
started. Human-team research has spent six decades asking exactly those questions, with
competing answers about where team phenomena live (in the members or in their interaction),
how team-level properties arise (by aggregation or by differentiation), and what distinguishes
what a team \emph{does} from how a team \emph{is}. I propose using that body of theory as the
organizing lens for configuring, observing and intervening in LLM-agent teams, without assuming
that its findings transfer.

I introduce \emph{\tk}, a framework and open reference implementation in which a user
configures a team of 1--16 AI agents for task completion, solution generation, achieving a goal
or a research mission;
selects one of fourteen deterministic routing topologies or a custom directed graph; chooses a
sharing profile that limits what may travel along permitted routes; and binds complementary
sources, external representations, work dependencies, version-specific handoffs and an
expertise directory. The same recorded process can then be observed through two groups of
views: \emph{non-system} views that locate explanation in member reports and behavior, and
\emph{system} views that locate it in distributed problem-solving, differentiated transactive
memory, interactive coordination and team dynamics. Interventions are explicit and human-led:
reviewed reconfiguration between runs, planned disturbances before a run, and private feedback,
pause, resume or stop applied at dependency-wave boundaries during a run. Every record keeps
``queued'', ``applied'', ``context assembled'', ``understood'' and ``improved'' as separate
claims. Emergent states such as trust, cohesion and shared understanding are declared unmeasured.

This is an idea and system-description paper. The contribution is a configuration space, a
reference implementation and a set of falsifiable hypotheses about topology, information flow
and intervention timing in LLM-agent teams. A planned line of study extends the author's earlier
work on the dynamics of distributed software teams: whether the fractal scaling and flexibility
of the communication output exchanged between members---and between members running on different
models---is associated with better task and team performance. No comparative result is reported,
and no topology is claimed to be best.
\end{abstract}

\section{Introduction}
\label{sec:intro}

A team of LLM agents has an organization whether or not anyone chose it. A chain of
specialists, a hub that collects proposals, a round of debate among peers and a fully
connected chat each decide who sees what and when. The choice is usually buried in the
architecture of the orchestration framework
\citep{li2023camel,hong2023metagpt,qian2023chatdev,wu2023autogen}, or delegated to an optimizer
that searches over communication graphs against a benchmark signal
\citep{zhuge2024gptswarm,zhang2025gdesigner,hu2024adas}. Both approaches answer the organizational
question before the work starts and give the user little to inspect or adjust once it has.

Human-team research has a long, unfinished argument about the same question. One tradition
treats the team as the context in which individual attributes add up: the input--process--output
(IPO) framework \citep{mcgrath1964social,hackman1987design} and the compositional reading of
team-level constructs \citep{kozlowski2000multilevel}. A second tradition moves to the system
level while keeping psychological constructs---emergent states such as shared mental models,
trust and psychological safety that develop over repeated episodes
\citep{marks2001temporal,ilgen2005imoi,edmondson1999safety}. A third treats team cognition as
something a team \emph{does} in interaction rather than something it \emph{has}
\citep{cooke2013interactive,gorman2017dynamical}, and reads variability in coordination as a
source of adaptability rather than noise. Gorman and colleagues \citep{gorman2017dynamical}
summarize the divide as non-system versus system explanations: whether a team phenomenon is
located in the members or in the interaction process.

The present paper takes these distinctions seriously as \emph{design questions for LLM-agent
teams}, while refusing the shortcut of assuming that human findings carry over. If the
organization of an agent team changes what it produces, a user should be able to choose the
organization explicitly, observe the work through more than one explanatory lens, and intervene
in ways whose scope is recorded. \tk{} is a concrete attempt to make that possible. It is a
follow-up to Labtune \citep{mitevski2026labtune}, which treated laboratory organization and
information sharing as controlled variables for LLM-agent research, and to the author's thesis
work on the dynamics of distributed software teams' creative processes \citep{mitevski2023thesis}.
Where Labtune offered six editable laboratory designs for research missions, \tk{} generalizes
the mission to tasks, brainstorming and goals, replaces lab designs with fourteen routing
topologies over 1--16 members, binds the team's problem-solving structure explicitly, and adds
research-informed views and human-led intervention. It ships its own runtime and makes no
automated organizational improvement.

The paper has five commitments:
\begin{itemize}
  \item \textbf{Configuration before execution.} Purpose, members, routing, sharing, bindings,
  models, tools and finite budgets are explicit and user-selected. Topology filters exposure;
  it grants no decision authority and creates no capacity (\cref{sec:model}).
  \item \textbf{Two explanatory lenses on one record.} Non-system and system views project
  the same configuration and trace. Switching views never changes the team (\cref{sec:views}).
  \item \textbf{Three distinctions kept apart.} Explanatory locus (non-system/system), form of
  emergence (compositional/compilational) and process/state/outcome are related but separate
  axes, and the implementation does not collapse them (\cref{sec:related}).
  \item \textbf{Explicit, bounded intervention.} Reviewed reconfiguration, planned disturbances
  and boundary-applied feedback, pause, resume and stop, with queued, applied, assembled,
  understood and improved recorded as distinct claims (\cref{sec:intervene}).
  \item \textbf{Declared non-measurement.} Trust, cohesion, psychological safety, efficacy,
  potency, shared understanding, emergence, synchrony and attractors are not inferred from
  software traces. What is recorded is operational; what is not measured is named.
\end{itemize}

\paragraph{Scope of this paper} This is an exploratory preprint and system description. I
contribute (i) the framing of team topology, information flow and intervention as explicit,
theory-indexed variables for LLM-agent teams; (ii) a formal configuration space and the design
principles that constrain it; (iii) an open reference implementation \citep{teamknitt2026code}
with offline demonstrations, grouped views and an intervention queue; and (iv) falsifiable
hypotheses and a protocol for testing them (\cref{sec:agenda}). I report no comparative result.
A persona label assigns a reasoning role; it is not evidence of expertise, and completion of a
workflow is not evidence that the user's goal was achieved.

\section{Background and related work}
\label{sec:related}

\subsection{Three distinctions from human-team research}
\label{sec:distinctions}

\paragraph{Explanatory locus} Non-system explanations locate a team phenomenon in member
attributes, knowledge and skills, or in their aggregate; system explanations locate it in the
interaction process, which has properties no member possesses alone \citep{gorman2017dynamical}.
This is an orientation, not a quality ranking, and system is not a synonym for complexity
science: conventional IPO and IMOI models also include interaction, feedback and contingencies
\citep{ilgen2005imoi}.

\paragraph{Form of emergence} Kozlowski and Klein \citep{kozlowski2000multilevel} distinguish
\emph{compositional} emergence, where a team-level construct is isomorphic with its member-level
counterpart and can be formed by means, sums or agreement, from \emph{compilational}
emergence, where the team-level pattern is a differentiated configuration with no simple
individual analogue. Shared mental models measured as within-team agreement
\citep{cannonbowers1993shared,mathieu2000shared,dechurch2010cognitive} are the textbook
compositional case; transactive memory---knowing who knows what, with differentiated encoding
and retrieval \citep{wegner1987transactive,lewis2003transactive}---is compilational. The
distinction is related to explanatory locus but not identical to it.

\paragraph{Process, state and outcome} Marks, Mathieu and Zaccaro \citep{marks2001temporal}
separate what teams do (transition and action processes, with interpersonal processes spanning
both) from emergent states that describe how teams are, and organize processes temporally
across recurring episodes. Outcomes feed later inputs \citep{ilgen2005imoi}. The distinction
matters for software: a publication, an exposure or an acknowledgement is an operational
process record. It is not cohesion \citep{beal2003cohesion}, trust, psychological safety
\citep{edmondson1999safety} or shared understanding.

\paragraph{Distributed, interactive and dynamical accounts} Distributed cognition
\citep{hutchins1995wild} and its extension to teams \citep{mcneese2021distributed} widen the
unit of analysis to people, artifacts, tools and representation transformations. Interactive
team cognition \citep{cooke2013interactive} treats cognition as situated interaction activity.
Coordination dynamics \citep{kelso1995dynamic,schmidt2008interpersonal} and the dynamical-systems
view of teams \citep{gorman2017dynamical} supply principles---local--global stability, substrate
independence, cross-level effects---and interventions such as perturbation training
\citep{gorman2010adaptive}, alongside meta-analytic evidence on team training
\citep{salas2008training}. Complex adaptive systems are a wider tradition
\citep{holland1995hidden,mitchell2009complexity}; citing them motivates questions, not
conclusions.

\subsection{Distributed teams and creative dynamics}
\label{sec:thesis}

The dynamical view also motivated the author's earlier empirical work on distributed software
teams \citep{mitevski2023thesis}. Using GitHub records of one organization, that study applied
detrended fluctuation analysis to code-addition time series and estimated maximal Lyapunov
exponents of daily discussion density, following Guastello's treatment of creative discussion
groups \citep{guastello1998chaos}. Long-range correlation was present in the development
dynamics, and the discussion that accompanied the more stable software product was more
flexible in the quantitative sense used there. The thesis also noted that such measurements
could inform when, and how much, an artificial teammate should intervene in a creative process.
\tk{} follows up on that interest from the other side: it records the team process at the
granularity of dependency waves and makes the intervention decision explicit and human-led.
It does \emph{not} estimate those exponents. Wave records are short and unevenly spaced; the
estimators belong in a research agenda (\cref{sec:agenda}), not in a product dashboard. The
agenda carries the question forward in a form the earlier data could not support: because every
contribution, exposure and acknowledgement in a \tk{} run is timestamped and attributed, the
communication output between members---its volume, its timing and its lexical content---can be
assembled into series long enough to ask whether its scaling properties relate to how well the
team did, and whether that relation changes when members run on different models.

\subsection{LLM-agent systems}
\label{sec:llmrelated}

\Cref{tab:related} places \tk{} among LLM-based approaches that touch the organizational
question. Reasoning methods such as Tree and Graph of Thoughts
\citep{yao2023tot,besta2024got} structure intermediate thoughts within an agent; \tk{} is
agnostic about them and organizes agents. Research systems---Agent Laboratory
\citep{schmidgall2025agentlab}, the AI co-scientist \citep{gottweis2025coscientist}, the AI
Scientist \citep{lu2024aiscientist}, the Virtual Lab \citep{swanson2025virtuallab}---fix a
division of labour and an exchange pattern; AgentRxiv \citep{schmidgall2025agentrxiv} adds
cross-lab sharing of reports. MacNet \citep{qian2024macnet}, GPTSwarm \citep{zhuge2024gptswarm}
and G-Designer \citep{zhang2025gdesigner} study or optimize communication topology; IMACS
\citep{chen2026imacs} routes tasks to protocols with a bandit; AutoGen \citep{wu2023autogen}
is general orchestration infrastructure. Failure analyses \citep{cemri2025fail}, evidence that
teams can hold experts back \citep{pappu2026experts}, and evidence that generative tools raise
individual output while reducing collective diversity \citep{doshi2024creativity} all suggest
that more communication is not automatically better collaboration. The Collaborative
Intelligence benchmark \citep{wu2026cib} evaluates an agent as a teammate on eight capabilities
after conditioning on solo competence; \tk{} treats that as a member-focused (non-system)
view and keeps probe evidence in a separate context.

\begin{table}[!tb]
\centering
\small
\caption{\tk{}, Labtune and related LLM-based approaches. The ``emphasis'' column states what
the present work contributes relative to each approach; it is not a claim of superiority.}
\label{tab:related}
\footnotesize
\begin{tabularx}{\linewidth}{@{}>{\raggedright\arraybackslash}p{2.9cm} >{\raggedright\arraybackslash}p{2.2cm} L L@{}}
\toprule
Approach & Category & Main focus & Labtune / \tk{} emphasis \\
\midrule
Tree / Graph of Thoughts \citep{yao2023tot,besta2024got} & Reasoning methods & Reasoning paths and dependencies between intermediate thoughts & Organization of agents and teams; reasoning methods can operate within individual members \\
Agent Laboratory \citep{schmidgall2025agentlab} & AI research system & Specialized agents covering literature review, experimentation and reporting & Alternative lab designs, topologies and exchange rules as user configurations \\
AgentRxiv \citep{schmidgall2025agentrxiv} & Research-sharing infrastructure & Shared reports so autonomous labs can build on each other's findings & Configurable exchange within and across teams and labs: who shares, when, under which rules and word budgets \\
AI co-scientist \citep{gottweis2025coscientist} & AI research system & Hypothesis generation, review and evolution through specialists and tournaments & Exploration of different organizations and exchange styles rather than one fixed process \\
MacNet \citep{qian2024macnet} & Multi-agent architecture & Graph-based collaboration; effects of topology and agent scaling & Topology combined with sharing profiles, bindings and user-led follow-up and intervention \\
GPTSwarm \citep{zhuge2024gptswarm} & Agent graph optimization & Optimization of prompts and information-flow connections & Explicit, stable configurations for examining how organizational choices affect results \\
G-Designer \citep{zhang2025gdesigner} & Topology optimization & Task-adaptive communication topologies balancing quality and token cost & Cross-domain comparison of topologies and exchange styles under controlled, non-growing budgets \\
AutoGen \citep{wu2023autogen} & Orchestration framework & General infrastructure for agent communication and orchestration & A team-specific framework centred on organization and its recorded effects \\
Collaborative Intelligence \citep{wu2026cib} & Teammate evaluation & Eight teammate capabilities, conditioned on solo competence & Member-focused view; separate-context probes, not an official score for a configured team \\
Labtune \citep{mitevski2026labtune} & Configurable research labs & Six lab designs, sharing levels, protected confirmation, multi-lab studies & Research missions and statistical confirmation \\
\textbf{\tk{}} (this paper) & Configurable task teams & Fourteen topologies, 1--16 members, sharing profiles, bindings, grouped views, interventions & Tasks, brainstorming, goals and research; human-team theory as lens \\
\bottomrule
\end{tabularx}
\end{table}

\paragraph{Relation to Labtune} Labtune \citep{mitevski2026labtune} asked which laboratory
organization, under which communication policy, suits which research problem, and supplied the
apparatus for asking it. \tk{} is based on Labtune and keeps its conservative core: budgets are
conserved under delegation, information crosses boundaries only through controller-published
views at phase barriers, and the system never redesigns its own organization. It departs in
four ways. The mission generalizes from research to any task, brainstorming prompt or goal.
The organizational variable changes from six prose-defined lab designs to fourteen deterministic
routing presets plus custom directed links, with explicit bindings of sources, representations,
work and handoffs. The observation layer is grouped by explanatory locus rather than by
research phase. And intervention becomes a first-class, bounded operation. Bayesian analysis
and model-assisted priors are not carried over; descriptive records and frequentist count and
interval analysis remain where applicable.

\section{Design principles}
\label{sec:principles}

The implementation follows eight rules. They are conservative on purpose: the aim is to make
organizational differences observable, not to make a team appear autonomous or effective.

\begin{enumerate}[label=\textbf{P\arabic*.},leftmargin=*]
  \item \textbf{The user configures the team.} Purpose, member count, roles, topology,
  sharing, models, tools, delegation and budgets are explicit. Drafts write nothing; applying a
  draft is a separate reviewed step with a private backup, blocked by the run lock. Larger
  teams never raise budgets.
  \item \textbf{Topology filters exposure and nothing else.} Routes decide which peer views
  and generated notes a member may encounter. A hub routes information; it acquires no veto.
  A hierarchy is a tree of routes, not a chain of command. Directed routes imply no reply.
  \item \textbf{Sharing is a policy applied at barriers.} Publications validate before a wave
  commits; unfinished same-wave output is never exposed. A read budget bounds cross-member
  exposure per episode, including generated notes.
  \item \textbf{Exposure is not noninterference.} A later contribution can relay what a member
  encountered along any later permitted route. Changing a route does not erase previous
  exposure; the record keeps the history.
  \item \textbf{Access, exposure, noticing, publication, citation and acknowledgement are
  distinct.} An acknowledgement confirms that a specific version was encountered; it does not
  imply agreement, understanding or correctness.
  \item \textbf{Views never change the team.} Non-system and system views, snapshots and the
  live monitor are presentation-only; they make no model calls and cannot write configuration.
  \item \textbf{Interventions are explicit, private and bounded.} Requests are drafts until
  applied, bind one run and configuration, and are acknowledged at the next dependency-wave
  boundary. Feedback is task context, never a permission or budget grant.
  \item \textbf{Nothing is inferred that is not measured.} Emergent psychological states,
  emergence itself, attractors, synchrony and causal intervention effects are not estimated
  from execution logs. Completion is not goal achievement.
\end{enumerate}

\section{The \tk{} model}
\label{sec:model}

\begin{figure}[!tb]
\centering
\begin{tikzpicture}[
  font=\scriptsize,
  stp/.style={draw, rounded corners=2pt, align=center, text width=3.05cm, minimum height=1.7cm, inner sep=3pt},
  sup/.style={draw, rounded corners=2pt, align=center, text width=3.05cm, minimum height=1.5cm, inner sep=3pt},
  arr/.style={-{Latex[length=1.6mm]}, thick},
  up/.style={-{Latex[length=1.6mm]}}
]
\node[stp, fill=KnittPaper] (s1) {\textbf{1 / Configure}\\[1pt] members, roles, topology, sharing;\\ explicit models, tools, budgets};
\node[stp, fill=KnittSand, right=0.3cm of s1] (s2) {\textbf{2 / Set tasks}\\[1pt] brief, sources, deliverable,\\ success criteria, review boundary};
\node[stp, fill=KnittMist, right=0.3cm of s2] (s3) {\textbf{3 / Observe work}\\[1pt] non-system and system views;\\ live metadata; offline snapshots};
\node[stp, fill=KnittSand, right=0.3cm of s3] (s4) {\textbf{4 / Intervene}\\[1pt] preview, then explicit apply;\\ human outcome review};
\foreach \a/\b in {s1/s2,s2/s3,s3/s4} \draw[arr] (\a) -- (\b);
\node[sup, below=1.0cm of s1, fill=KnittOchre!14] (b1) {\textbf{Configuration identity}\\ drafts write nothing; apply backs up;\\ no automatic redesign};
\node[sup, below=1.0cm of s2, fill=KnittOchre!14] (b2) {\textbf{Barrier publication}\\ validate, then expose along routes;\\ read budget $W$ per episode};
\node[sup, below=1.0cm of s3, fill=KnittMoss!12] (b3) {\textbf{Operational records}\\ available / exposed / published /\\ acknowledged; states unmeasured};
\node[sup, below=1.0cm of s4, fill=KnittMoss!12] (b4) {\textbf{Bounded queue}\\ queued $\neq$ applied $\neq$ assembled\\ $\neq$ understood $\neq$ improved};
\foreach \a/\b in {b1/s1,b2/s2,b3/s3,b4/s4} \draw[up] (\a) -- (\b);
\draw[arr, KnittInk!65] (s4.north) -- ++(0,0.4) -| node[pos=0.25, above, font=\scriptsize\itshape]{between runs: reviewed reconfiguration} (s1.north);
\end{tikzpicture}
\caption{\tk{} from configuration to reviewed output. Top row: what the user does. Bottom row:
the records that govern each step. Reconfiguration happens between runs; views and interventions
never alter the organization of a run in progress.}
\label{fig:overview}
\end{figure}
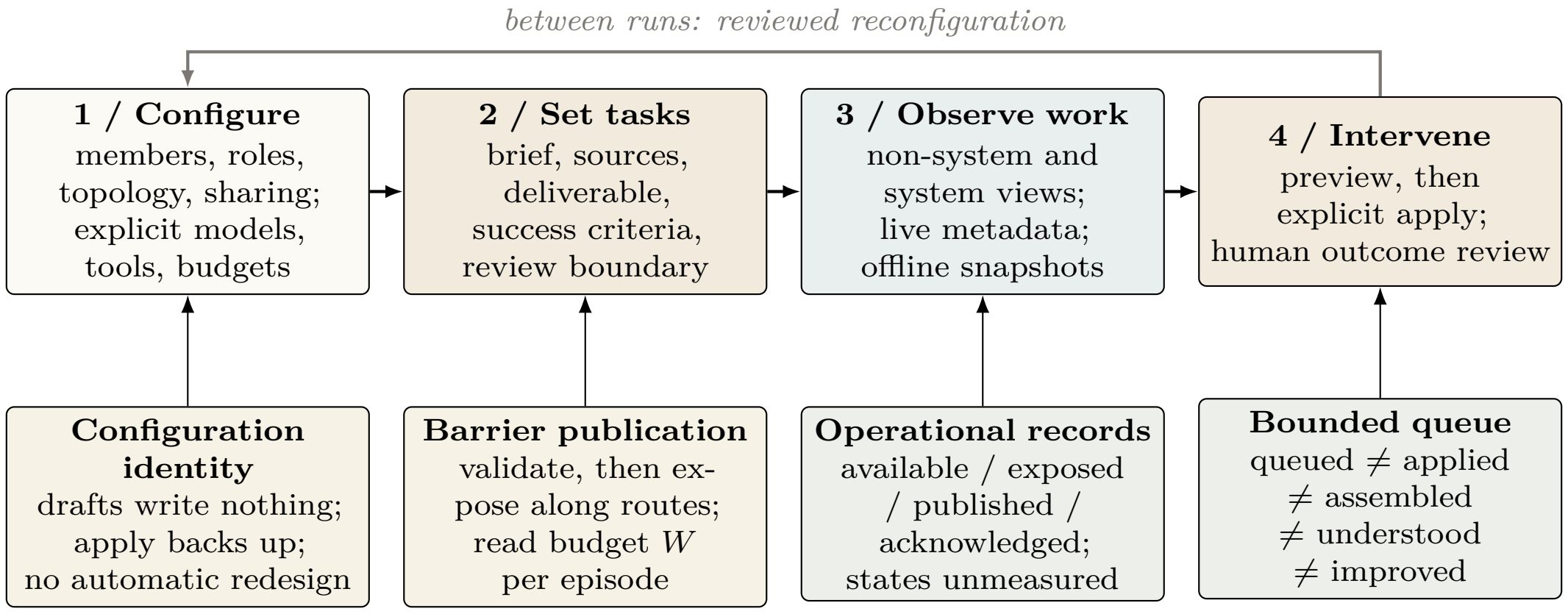

\Cref{fig:overview} shows the four steps the user moves through and the records that govern
each step. The user configures a team, sets the task, observes the work through chosen views,
and intervenes within recorded scope. The engine compiles the configuration into a task-team
workflow of dependency waves, schedules bounded member frames, publishes permitted views at
barriers, and retains contributions for human review.

\subsection{Configuration space}
\label{sec:config}

A team configuration is a tuple
\begin{equation}
  \mathcal{T} = (p,\; N,\; G,\; \pi,\; \Sigma,\; \Delta,\; \mathcal{M},\; \mathcal{B},\; \mathcal{D}),
\end{equation}
where $p$ is the purpose, one of \textsf{task}, \textsf{brainstorm}, \textsf{goal} or
\textsf{research}; $N \in \{1,\dots,16\}$ the member count; $G = (V, E)$ a directed routing graph over
members $V = \{1,\dots,N\}$ drawn from a preset or given as custom links (\cref{sec:topology});
$\pi = (\ell, w, W)$ the sharing policy, with profile $\ell$ one of \textsf{none},
\textsf{low}, \textsf{intermediate}, \textsf{high} or \textsf{wisdom\_only}, wisdom flag $w$ and
read budget $W$ in words (\cref{sec:sharing}); $\Sigma$ the problem-solving
bindings (\cref{sec:bindings}); $\Delta$ the delegation limits; $\mathcal{M}$ the provider,
model and tool bindings; $\mathcal{B}$ the finite budgets (requests, wall time, concurrency);
and $\mathcal{D}$ an optional set of planned disturbances (\cref{sec:intervene}). The
configuration is loaded from strict YAML, validated, and hashed into an identity stored with
every plan, run and snapshot. Initialization, drafting, planning, viewing and demonstration
make no provider calls, so the configuration space can be explored for free.

\subsection{Routing topologies}
\label{sec:topology}

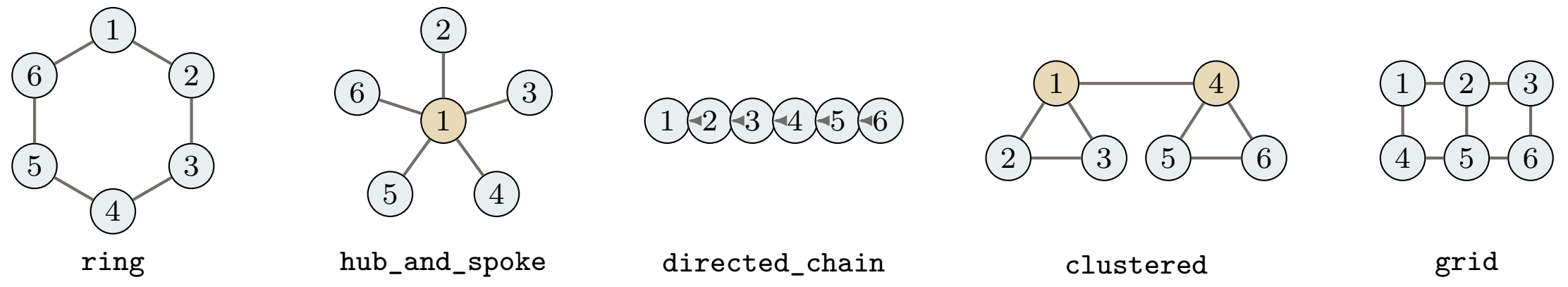
\begin{figure}[!tb]
\centering
\begin{tikzpicture}[
  font=\scriptsize,
  m/.style={circle, draw, fill=KnittMist, minimum size=4.2mm, inner sep=0pt},
  h/.style={circle, draw, fill=KnittOchre!40, minimum size=4.2mm, inner sep=0pt},
  rec/.style={thick, KnittInk!70},
  dir/.style={thick, KnittInk!70, -{Latex[length=1.4mm]}}
]
\begin{scope}[xshift=0cm]
\foreach \i in {1,...,6} \node[m] (r\i) at ({90-60*(\i-1)}:0.85) {\i};
\foreach \i/\j in {1/2,2/3,3/4,4/5,5/6,6/1} \draw[rec] (r\i) -- (r\j);
\node at (0,-1.35) {\code{ring}};
\end{scope}
\begin{scope}[xshift=3.1cm]
\node[h] (c) at (0,0) {1};
\foreach \i in {2,...,6} \node[m] (s\i) at ({90-72*(\i-2)}:0.85) {\i};
\foreach \i in {2,...,6} \draw[rec] (c) -- (s\i);
\node at (0,-1.35) {\code{hub\_and\_spoke}};
\end{scope}
\begin{scope}[xshift=6.2cm]
\foreach \i in {1,...,6} \node[m] (d\i) at ({-1.0+0.4*(\i-1)},0) {\i};
\foreach \i/\j in {1/2,2/3,3/4,4/5,5/6} \draw[dir] (d\i) -- (d\j);
\node at (0,-1.35) {\code{directed\_chain}};
\end{scope}
\begin{scope}[xshift=9.6cm]
\node[h] (c1) at (-0.75,0.35) {1}; \node[m] (c2) at (-1.2,-0.35) {2}; \node[m] (c3) at (-0.3,-0.35) {3};
\node[h] (c4) at (0.75,0.35) {4}; \node[m] (c5) at (0.3,-0.35) {5}; \node[m] (c6) at (1.2,-0.35) {6};
\foreach \i/\j in {1/2,2/3,1/3,4/5,5/6,4/6} \draw[rec] (c\i) -- (c\j);
\draw[rec] (c1) -- (c4);
\node at (0,-1.35) {\code{clustered}};
\end{scope}
\begin{scope}[xshift=12.7cm]
\foreach \i in {1,2,3} \node[m] (g\i) at ({-0.6+0.6*(\i-1)},0.35) {\i};
\foreach \i in {4,5,6} \node[m] (g\i) at ({-0.6+0.6*(\i-4)},-0.35) {\i};
\foreach \i/\j in {1/2,2/3,4/5,5/6,1/4,2/5,3/6} \draw[rec] (g\i) -- (g\j);
\node at (0,-1.35) {\code{grid}};
\end{scope}
\end{tikzpicture}
\caption{Five of the fourteen presets for $N=6$. Plain lines are reciprocal routes; arrows are
one-way routes. Shaded members occupy routing positions (hub, pod lead), which carry no decision
authority.}
\label{fig:topologies}
\end{figure}

Fourteen presets define $E$ deterministically for any $N \le 16$ (\cref{tab:topologies};
\cref{fig:topologies}). Reciprocal presets add both directions of a link; directed presets add
one. A single member has no peer links, and small layouts can coincide (a two-member ring is a
chain). Custom links are exactly the listed directed pairs; an empty list isolates every
member. Presets are routing definitions, not inferred optimal organizations. Two consequences
follow from P2 and P4. First, member 1 in a hub, wheel, broadcast or gather layout is a
routing position, not a leader: it cannot veto, rank or create resources. Second, a diagram
line denotes a permitted route, not causal influence; with sufficient waves, content can relay
across a chain or around a ring, and the record will show where it did.

\begin{table}[!tb]
\centering
\small
\caption{The fourteen routing presets. All support 1--16 members.}
\label{tab:topologies}
\begin{tabularx}{\linewidth}{@{}l L@{}}
\toprule
Preset & Exact routing \\
\midrule
\code{fully\_connected} & Every ordered pair of distinct members \\
\code{hub\_and\_spoke} & Member 1 exchanges two-way with every spoke; no spoke--spoke links \\
\code{ring} & Two-way neighbours with a closing loop \\
\code{hierarchical} & Two-way binary tree over member ids; no decision authority is granted \\
\code{chain} & Two-way adjacent ids, no closing loop \\
\code{directed\_chain} & $1 \to 2 \to \dots \to N$; no implied replies \\
\code{directed\_ring} & Each member sends to its successor; $N$ sends to 1 \\
\code{wheel} & Two-way member-1 hub plus a two-way ring among the spokes \\
\code{grid} & Row-major $\lceil\sqrt{N}\rceil$ columns; horizontal and vertical neighbours, no wrap or diagonals \\
\code{clustered} & Consecutive pods of up to three, fully connected within a pod; successive pod leads connect two-way \\
\code{bipartite} & First $\lceil N/2 \rceil$ members connect two-way to all remaining members; no links within a partition \\
\code{broadcast} & Member 1 sends to all others; no replies or spoke links \\
\code{gather} & All others send to member 1; no return or spoke links \\
\code{custom} & Exactly the listed directed source/target pairs; empty means isolated \\
\bottomrule
\end{tabularx}
\end{table}

\subsection{Sharing profiles and exposure}
\label{sec:sharing}

Sharing limits how much, and what kind of, content can travel along permitted routes
(\cref{tab:sharing}). Let $o_i(t)$ denote member $i$'s published representations at barrier
$t$, and $V_\pi$ the controller's filter for profile $\pi$. Member $j$'s context evolves as
\begin{equation}
  C_j(t+1) \;=\; C_j(t) \,\cup\, \operatorname{trunc}_W\!\Big(V_{\pi}\big(\{\,o_i(t) : (i \to j) \in E\,\}\big)\Big),
\end{equation}
where $\operatorname{trunc}_W$ bounds the combined exposure, including generated notes, to $W$
words per episode. Notes are not an extra channel: they obey the same routes and the same
budget. User-supplied sources have their own reader grants and are not subject to $E$.
Shared writable workspaces require \textsf{high}. All publications in a wave validate before
the wave commits, so a member never sees a peer's unfinished same-wave output. The equation
describes a filter on exposure, not a guarantee of noninterference (P4): what $j$ encounters at
$t$ may reappear, transformed, in $o_j(t+1)$ and travel further along $(j \to k) \in E$.

\begin{table}[!tb]
\centering
\small
\caption{Sharing profiles. A profile limits content along permitted routes; it does not add
routes.}
\label{tab:sharing}
\begin{tabularx}{\linewidth}{@{}>{\raggedright\arraybackslash}p{2.4cm} L L@{}}
\toprule
Profile & Admitted content & Intended trade-off \\
\midrule
\textsf{none} & No peer content; independent contributions retained for user review & Baseline that preserves independent search; joint phases are operational only \\
\textsf{low} & Bounded member-reported blockers, labelled unverified & Avoid repeated dead ends without steering toward a solution \\
\textsf{intermediate} & Compact contribution summaries & Share direction while limiting copying \\
\textsf{high} & Detailed peer views and explicitly granted versioned notes & Deliberate reuse and recombination \\
\textsf{wisdom\_only} & Conditional lessons with stated limits, not detailed peer output & General lessons with reduced solution steering \\
\bottomrule
\end{tabularx}
\end{table}

\subsection{Problem-solving bindings}
\label{sec:bindings}

The bindings $\Sigma$ make the team's problem-solving structure explicit so that system views
have something to inspect other than chat. $\Sigma$ declares: \emph{roles} with
responsibilities, a perspective and capabilities; \emph{representations}---user sources and
per-member append-only structured notes---with explicit readers and writers;
\emph{work units} with phase, actors, inputs, outputs and dependencies; version-specific
\emph{handoffs} from a producer to a consumer with an acknowledgement requirement; an
\emph{expertise directory} that records which members are assigned which topics, with the
declared basis (``user-editable role emphasis, not verified expertise''); a \emph{temporal
reference} ordering the phases with a deadline wave; and \emph{affordances}, the capability-conditioned
action opportunities a member actually has (publish to a note; acknowledge a
handoff). \code{teamknitt init} generates these bindings mechanically from purpose, member
count, topology and sharing; the user edits them directly rather than regenerating. The default
task-team design runs four phases---independent proposals, task adoption, parallel episodes
and evidence review---and asks members to frame, connect, develop and review while preserving
alternatives before converging.

\subsection{Execution loop}
\label{sec:loop}

\begin{figure}[!tb]
\centering
\begin{tikzpicture}[
  font=\scriptsize,
  b/.style={draw, rounded corners=2pt, align=center, text width=2.55cm, minimum height=1.25cm, inner sep=3pt},
  arr/.style={-{Latex[length=1.6mm]}, thick},
  darr/.style={-{Latex[length=1.6mm]}, dashed}
]
\node[b, fill=KnittPaper] (src) {\textbf{User brief and sources}\\ task, prompt or goal;\\ complementary inputs};
\node[b, fill=KnittMist, right=0.45cm of src] (act) {\textbf{Member's permitted view}\\ own role, representations,\\ affordances};
\node[b, fill=KnittMoss!14, right=0.45cm of act] (pub) {\textbf{Externalize contribution}\\ bounded note with\\ input provenance};
\node[b, fill=KnittMoss!14, right=0.45cm of pub] (bar) {\textbf{Wave barrier}\\ validate all publications;\\ commit immutable versions};
\node[b, fill=KnittMist, right=0.45cm of bar] (con) {\textbf{Peer's later view}\\ exposure along $E$ under $\pi$;\\ version-specific ack};
\node[b, fill=KnittSand, below=0.7cm of pub] (ctl) {\textbf{Team choices + bounds}\\ $G$, $\pi$, $W$, models, tools,\\ budgets; intervention queue};
\node[b, fill=KnittOchre!16, below=0.7cm of con] (rec) {\textbf{Process records}\\ available / exposed / published /\\ acknowledged; freshness};
\draw[arr] (src) -- (act);
\draw[arr] (act) -- (pub);
\draw[arr] (pub) -- (bar);
\draw[arr] (bar) -- (con);
\draw[arr] (ctl) -- (bar);
\draw[arr] (bar) |- (rec);
\draw[darr] (con.north) -- ++(0,0.45) -| node[pos=0.25, above, font=\scriptsize\itshape]{next wave} (act.north);
\end{tikzpicture}
\caption{One dependency wave. Routes $E$ and profile $\pi$ filter what a peer may encounter;
the barrier publishes only validated, immutable versions; the intervention queue is read before
a wave begins, outside any publication transaction.}
\label{fig:protocol}
\end{figure}
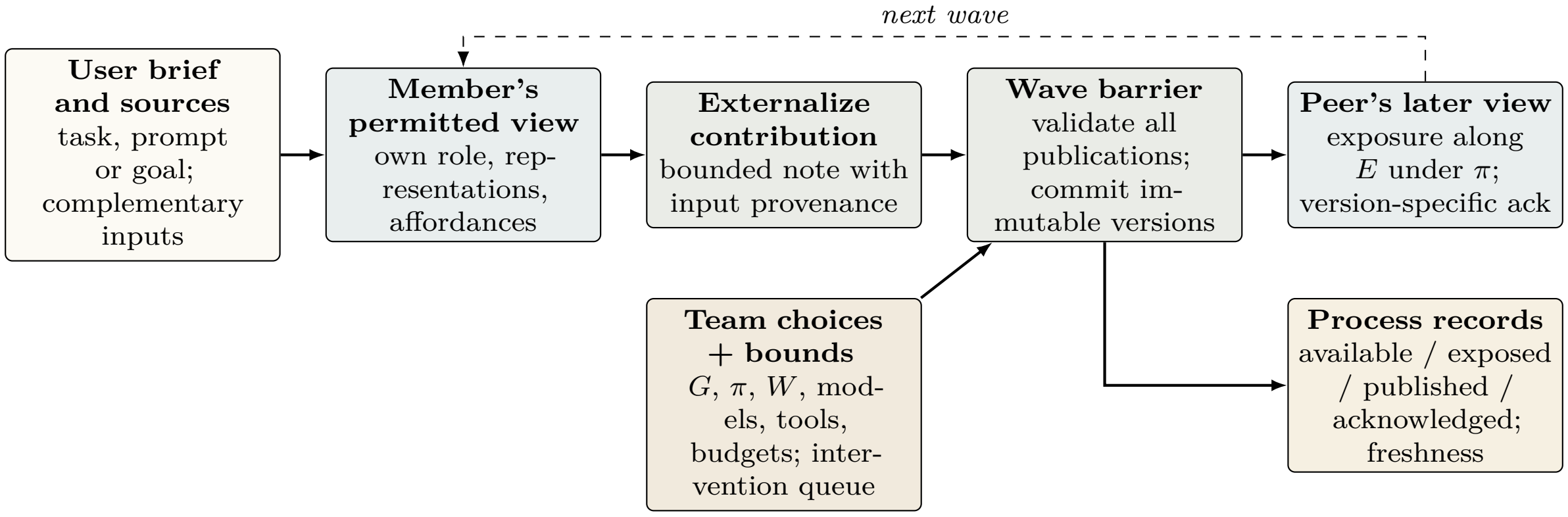

\Cref{fig:protocol} shows one dependency wave. A member observes only its permitted inputs,
representations and affordances, externalizes a bounded contribution with input provenance,
and the controller validates all publications of the wave before committing immutable versions.
Consumers encounter permitted versions in a later wave and acknowledge specific handoffs. The
record distinguishes available, exposed, published and acknowledged, and keeps report freshness
and missingness explicit. Delegation, decision tools, retries and corrections draw on the same
request cap. The default contribution is bounded text; code or external actions require an
explicitly permitted tool and its receipt, and the design instructs members not to pretend
otherwise. Completion means contributions were produced and retained for review in
\code{trials/ID/deliverable.md}; it is not a verified outcome.

\section{Observing the work: two groups of views}
\label{sec:views}

The same configuration and record can be observed through two groups of views
(\cref{tab:views}). The grouping follows explanatory locus (\cref{sec:distinctions}). It is an
orientation for the observer, not a partition of science or a pair of execution presets, and
switching groups is presentation-only.

\begin{table}[!tb]
\centering
\small
\caption{Grouped views over one configuration and record. The right-hand column names what the
view does \emph{not} measure.}
\label{tab:views}
\footnotesize
\begin{tabularx}{\linewidth}{@{}>{\raggedright\arraybackslash}p{3.0cm} L L@{}}
\toprule
View & Inspects & Not measured or inferred \\
\midrule
\multicolumn{3}{@{}l}{\textit{Non-system: member- and component-centred}} \\
Collaborative intelligence & Eight teammate-behaviour questions \citep{wu2026cib}, from an optional, separately run probe protocol & Capabilities of the configured team; an official benchmark score; psychological traits \\
Shared models \& member reports & Explicit plans, agreement with a declared reference, report freshness, perceived alignment & Latent mental-model similarity or accuracy; understanding; correctness \\
\addlinespace
\multicolumn{3}{@{}l}{\textit{System: interaction- and configuration-centred}} \\
Distributed problem-solving & Complementary inputs, representations, tools, expertise directory, transformations, dependencies, handoffs & Emergent cognitive capacity; goal achievement \\
Transactive memory & Differentiated (compilational) expertise assignments, holders and retrieval scaffolds & Credibility, specialization, retrieval success, system effectiveness \\
Interactive cognition \& coordination & Observed artifact exposure, version-specific handoff receipts, cross-level records & Understanding or agreement; configured routes are never substituted for missing observations \\
Team dynamics & Selected temporal activity, continuity, disturbance windows, return and maintenance, first attainment, censoring & Attractors, synchrony, fractal scaling, emergence, adaptation, causality \\
\bottomrule
\end{tabularx}
\end{table}

\paragraph{Non-system views} The collaborative-intelligence view is member-focused and
behavioural. \tk{} ships original synthetic alone/baseline/challenged probes informed by the
Collaborative Intelligence benchmark \citep{wu2026cib}; they run in a separate workspace, use
alone performance for eligibility, leave incomplete cells unknown rather than zero, and can be
attached to a task view only as separate-context evidence. Ordinary task telemetry does not
evaluate the eight capabilities. The shared-models view compares explicit member reports:
reported plans, conformity to a declared reference, freshness and perceived alignment. These
are compositional proxies by construction and are labelled as such; similarity is not accuracy
\citep{mathieu2000shared}.

\paragraph{System views} The distributed problem-solving view inspects the bindings of
\cref{sec:bindings} together with recorded transformations and handoffs, in the spirit of
\citep{hutchins1995wild,mcneese2021distributed}. The transactive-memory view inspects the
configured expertise directory as a differentiated pattern
\citep{wegner1987transactive,lewis2003transactive}, and says so when the directory is merely
configured rather than demonstrated. The interactive-cognition view inspects exposure and
acknowledgement receipts \citep{cooke2013interactive}; a configured route is never substituted
for a missing observation. The dynamics view exposes optional descriptive observations.

\paragraph{Optional process observations} When enabled, four descriptive dimensions are
derived from the record: \emph{local--global}, contribution and receipt activity alongside
handoff-state continuity; \emph{perturbation}, explicit disturbance windows with baseline
sufficiency, overlap, maintenance or return, first attainment and right censoring;
\emph{coupling}, directed artifact-exposure and provenance counts; and \emph{cross-level},
member, artifact and coordination records together. Three further dimensions---attractors,
synchronization and fractals---are listed so that their absence is explicit: no estimator is
implemented, and dependency waves are neither evenly spaced nor uniform in content. Variability
in contribution activity is reported, not read as flexibility or resilience.

\section{Intervention}
\label{sec:intervene}

Human-team research addresses intervention at several levels: staffing and role design, task
design, procedures and enabling conditions \citep{hackman1987design,hackman2002leading};
cross-training and perturbation training \citep{gorman2010adaptive}; and team training in
general \citep{salas2008training}. \tk{} offers one explicit mechanism for each timing
(\cref{tab:interventions}) and refuses to collapse them: a changed input, a changed process, a
report of alignment and an improved outcome are different claims.

\begin{table}[!tb]
\centering
\small
\caption{Intervention mechanisms by timing. Human findings motivate the mechanisms; they do not
establish effects for AI teams.}
\label{tab:interventions}
\footnotesize
\begin{tabularx}{\linewidth}{@{}>{\raggedright\arraybackslash}p{3.2cm} L L@{}}
\toprule
Timing / mechanism & Actual support & Boundary \\
\midrule
Between runs: inputs and conditions & Reviewed changes to roles, task, context, criteria, sources, procedures, topology and sharing & Draft first; explicit apply with private backup under the run lock; no automatic staffing or model choice \\
Before a run: planned disturbance & \code{actor\_unavailable}, \code{representation\_unavailable} or \code{link\_dropout} at a chosen wave for a chosen duration & A software condition for a future run, not weight training or a validated adaptive-training replication \\
During a run: private feedback & Bounded task context for chosen members' future frames; newest text first within the representation word cap & No new tool, source or permission grants; not proof of delivery or understanding \\
During a run: pause / resume / stop & Acknowledged at the next dependency-wave boundary; current-wave work may finish & Pause keeps the lock, counters and wall deadline; resume resets nothing; stop is terminal and skips confirmation \\
After a run: outcome review & Private partial or complete deliverable, request receipts, separate outcome evidence & No automatic quality, psychological-state or intervention-effect score \\
\bottomrule
\end{tabularx}
\end{table}

\paragraph{Queue semantics} Requests are prepared from the command line (the browser desk
prepares commands and reads receipts; it cannot write the queue or start execution). A request
without \code{--apply} is a preview. An applied request binds one run and one configuration
identity and is read by the runtime before it begins a dependency wave, outside any publication
transaction. The record then separates five claims that are easy to conflate:
\begin{equation*}
  \textit{queued} \;\neq\; \textit{applied at boundary} \;\neq\; \textit{context assembled}
  \;\neq\; \textit{understood} \;\neq\; \textit{improved}.
\end{equation*}
A context-assembly event is metadata about constructing a member's frame; it is not proof that
a provider received the text or that a model used it. A request arriving after the last useful
boundary is retained as not applied. Receipts carry recipients, timestamps, hashes, boundary
waves and assembly counts, never message bodies. The queue is bounded (256 requests per run,
64 pending) and is a private file queue, not a multi-user or cryptographic authorization system.

\paragraph{Planned disturbances} A disturbance is scheduled before a run, with an explicit
kind, target, wave and duration that must fit the configured system. It changes the
configuration identity of a future run and never a historical record. The perturbation
observation then reports baseline sufficiency, the disturbed window, whether a continuity
measure was maintained or returned, first attainment and censoring. It reports no recovery
score and infers no causal effect; the comparison condition, independent outcome evidence and
analysis are the user's.

\section{Reference implementation}
\label{sec:impl}

\tk{} 0.9.0 is a Python ($\geq$3.11) package with one console command and its own runtime,
\code{teamknitt.engine}, which owns scheduling, provider and decision adapters, persistence,
structural planning, schemas and resources \citep{teamknitt2026code}. It imports no upstream
package and delegates to no other CLI. The core modules comprise roughly 7{,}300 lines; the
offline test suite contains 21 modules and about 350 tests, and setup, planning, viewing,
drafting, demonstration, diagram rendering and packaging make no provider calls.
\Cref{tab:impl} separates implemented components from limits.

\begin{table}[!tb]
\centering
\small
\caption{Implementation status of \tk{} 0.9.0.}
\label{tab:impl}
\footnotesize
\begin{tabularx}{\linewidth}{@{}l L L@{}}
\toprule
Layer & Implemented & Not implemented / limits \\
\midrule
Configuration & Four purposes; 1--16 members; 14 presets plus custom links; five sharing profiles; schema-3 bindings; drafts, backups, run lock & No automatic topology, staffing or model selection \\
Runtime & Dependency waves, scoped frames, immutable versioned notes, route-filtered views, barrier publication, handoffs, plan reports & Bounded text by default; live mid-run reconfiguration; durable crash recovery \\
Views & Non-system and system groups; live metadata monitor; private offline snapshots; optional probe attachment & Metadata only, never raw prompts, generated text or credentials \\
Observations & Local--global, perturbation, coupling and cross-level descriptions & Attractors, synchronization and fractals not estimated; no emergence or causal inference \\
Interventions & Private file queue; feedback, pause, resume, stop at wave boundaries; planned disturbances & Not available to advanced engine workflows implicitly; no retries or recovery \\
Statistics & Descriptive records; frequentist count and interval analysis where applicable & Bayesian analysis and prior elicitation removed; unsupported legacy choices fail before provider initialization \\
Probes & Synthetic alone/baseline/challenged teammate probes in a separate workspace & Not the official Collaborative Intelligence benchmark or a score for a configured team \\
Advanced engine & Provider and decision adapters, research adapters, single-scientist and multi-lab workflows inherited from Labtune & Candidate execution sandbox limited to one synthetic domain \\
\bottomrule
\end{tabularx}
\end{table}

\paragraph{Offline demonstration} \code{teamknitt demo} runs deterministic ``connected'' and
``fragmented'' fixtures through the full configure--publish--observe path without a model, so
that routes, barriers, receipts and views can be inspected before any budget is spent. Fixtures
are software behaviour, not model evidence or a solution to the user's task. Live execution
requires an explicit provider, a private key, a reviewed plan and
\code{run --live --approve-requests N}, where $N$ must equal the resolved request cap.

\paragraph{Observability} A local, read-only viewer and the offline snapshot show the resolved
team, routes, bindings, grouped views, heartbeat and recorded completion. Live activity, stale
metadata and recorded completion are visually distinct. Nothing in the browser applies a
configuration, changes a budget or starts a model.

\section{Research agenda: hypotheses and protocol}
\label{sec:agenda}

\tk{} is useful only if organizational and intervention claims about LLM-agent teams become
testable. The hypotheses below are possibilities, not expectations; several are phrased so that
either outcome would be informative about whether human-team regularities transfer.

\begin{enumerate}[label=\textbf{H\arabic*.},leftmargin=*]
  \item \textbf{Topology is inert without sharing.} Under \textsf{none}, differences between
  presets in the diversity of retained alternatives vanish; they appear only at
  \textsf{intermediate} or above. (A manipulation check on P2--P3.)
  \item \textbf{Hubs converge faster, not better.} At matched $N$, $W$ and budget,
  \code{hub\_and\_spoke} retains fewer distinct alternatives at review than \code{ring} or
  \code{grid}, without a higher share of contributions judged useful by the user.
  \item \textbf{Direction limits relay.} \code{directed\_chain} preserves more of the
  first-wave alternatives than \code{chain} at equal exposure budget, because reciprocal routes
  allow content to return and recombine.
  \item \textbf{Redundancy and disturbance.} After an \code{actor\_unavailable} disturbance,
  handoff continuity returns sooner in topologies with alternative routes (\code{ring},
  \code{grid}, \code{wheel}) than in \code{chain} or \code{hub\_and\_spoke}; whether variability
  in contribution activity before the disturbance predicts return, as the dynamical view of
  human teams suggests \citep{gorman2017dynamical}, is an open question.
  \item \textbf{Timing of feedback.} Private feedback assembled before a transition phase
  (task adoption, evidence review) changes subsequent contributions more than feedback assembled
  during an action phase (parallel episodes), consistent with the temporal taxonomy of
  \citep{marks2001temporal}.
  \item \textbf{Member measures do not predict system records.} A member's separately measured
  teammate profile \citep{wu2026cib} predicts its in-team acknowledgement and relay behaviour
  only weakly once topology and sharing are controlled. A strong prediction would favour
  non-system explanation; a weak one would favour system explanation.
  \item \textbf{Same-model homogeneity.} When all members use one model, topology differences
  shrink as sharing increases; heterogeneous model assignment widens them.
  \item \textbf{Fractality of communication output.} Following
  \citep{mitevski2023thesis,guastello1998chaos}, the series of communication output exchanged
  between members (per-wave contribution and acknowledgement volume, inter-contribution
  intervals, and turnover in lexical content across versions) exhibits long-range correlation,
  and teams whose series show a scaling exponent in the persistent range ($0.5 < \alpha < 1$
  under detrended fluctuation analysis) and a positive but bounded maximal Lyapunov exponent
  for discussion density receive higher user-judged usefulness and retain more distinct
  alternatives than teams whose series are closer to white noise or to rigid periodicity. Scaling
  that reflects only the scheduler's wave structure must be ruled out with surrogate series and
  with the \textsf{none} baseline, where no inter-member communication occurs.
  \item \textbf{Cross-model fractality.} When members run on different models, the output of
  each model carries its own scaling signature, and the degree to which those signatures align
  after exchange (relative to runs where the same models never see each other's output) is
  associated with team performance; the sign of that association is left open, since alignment
  may indicate productive coupling or premature convergence.
\end{enumerate}

\paragraph{Protocol} I propose a factorial design over topology (a selected subset of the
fourteen presets), sharing profile (five) and member count (for example $N \in \{3, 6, 9\}$),
with purpose, brief, model, request cap, read budget $W$ and review criteria held fixed, and
\code{custom} with empty links as the isolated baseline. Each cell is replicated with fresh
contexts. Outcomes are operational and human-reviewed: the number of distinct alternatives
retained at review; the share of contributions the user judges useful against declared
criteria; relay distance of first-wave content; acknowledgement rate and latency; handoff
continuity around disturbances; and realized cost including corrections and delegation.
Analysis is descriptive with frequentist count and interval estimates declared before review.
\tk{}'s records---configuration identities, exposure logs, receipts and assembly events---are
designed to supply these quantities directly. H8 and H9 need longer runs than the other
hypotheses: scaling estimates require series of at least a few hundred points, so these studies
would use goal-oriented missions with many waves, or many short runs of the same configuration
whose per-wave series are analysed as an ensemble, with DFA and Lyapunov estimates compared
against phase-randomized and shuffled surrogates \citep{mitevski2023thesis}. The communication
series themselves would be derived offline from the exported event store; no estimator runs
inside the product.

\paragraph{Threats to validity} Four are immediate. Personas on one model may differ in label
more than in behaviour; the mechanisms must be measured, not assumed. Field-based sharing
filters cannot guarantee semantic isolation, so leakage is an outcome to quantify rather than a
property to assume. Provider nondeterminism and silent model changes require replication and
recorded model identities. Finally, human review of usefulness is itself a measurement with
reliability that must be checked.

\section{Limitations}
\label{sec:limitations}

The main limitation is the boundary of the paper: no empirical comparison of topologies,
sharing profiles or interventions is reported. \tk{} supplies the apparatus, not evidence that
any configuration is better. The grouped views are explanatory orientations applied to
operational scaffolds; they do not measure the psychological constructs whose names they
borrow, and the implementation says so at each view. The dynamical observations are descriptive,
and the estimators used in the author's earlier work \citep{mitevski2023thesis}---detrended
fluctuation analysis and Lyapunov exponents---are not applied inside the product, because
wave-level records from a single run are short and unevenly spaced; H8 and H9 therefore depend on
an offline analysis pipeline and on longer or pooled runs that do not yet exist. The teammate probes are original synthetic protocols informed by the
Collaborative Intelligence benchmark, not the benchmark itself. Intervention receipts record
assembly, not understanding; feedback effects are not estimated. The runtime defaults to bounded
text work; code and external actions require explicitly permitted tools. Finally, the software
is distributed under a source-available noncommercial and research licence rather than an OSI
open-source licence: general noncommercial use is free, but research use outside qualifying
academic institutions, and any commercial use, require separate written permission.

\section{Conclusion}
\label{sec:conclusion}

Every LLM-agent team has a topology, an information policy and a stance on human intervention,
even when the framework hides them. \tk{} makes those three choices explicit and user-owned,
and offers human-team theory as a lens for observing what happens rather than as a source of
borrowed conclusions. The user chooses who is connected to whom and what may travel; observes the
work through member-centred or interaction-centred views that never alter it; and intervenes
through mechanisms whose scope, timing and receipts are recorded. Configuration identities,
exposure logs and the chain from queued to improved then make it possible to ask, for agent
teams, the questions that human-team research has been asking for decades: \emph{where does a
team's success or failure live, how does a team-level property arise, and when does intervening
help?} Whether the answers for LLM agents resemble the answers for people is an empirical
question. \tk{} is intended to provide the apparatus for asking it.

\section*{Code availability}
Source code, topology catalog, task-team design, starters and diagrams are available (version
0.9.0) at \url{https://github.com/JeranRolevs/teamknitt}. \tk{} is based on Labtune
(\url{https://github.com/JeranRolevs/labtune}). The software is distributed under the Teamknitt
Noncommercial and Research Licence 2.0 (copyright holder and licensor: Pioneertown B.V.; author:
Goran Mitevski); see \code{LICENSE} and \code{LICENSING.md} in the repository for terms.
Credentials, trials and research results are not distributed.

\section*{Disclosure}
During the preparation of this manuscript, the author utilized a combination of LLM models to
rephrase and improve the clarity, readability and stylistic tone of the draft. After using these
tools, the author reviewed, edited and verified the content to ensure accuracy, and takes full
responsibility for the final manuscript.

\bibliographystyle{plainnat}
\bibliography{references}

\end{document}